\pdfoutput=1
\documentclass[twocolumn,superscriptaddress,aps,preprintnumbers,
prd,nofootinbib
]{revtex4-2}

\usepackage{graphicx}
\usepackage{xcolor}
\usepackage[caption=false]{subfig}
\usepackage{mathrsfs,mathtools}
\usepackage{physics,amssymb}
\usepackage{siunitx}
\usepackage{bm}
\usepackage{braket}
\usepackage{listings}
\usepackage{cases}
\usepackage{comment}
\usepackage{soul}
\usepackage{cancel}
\usepackage{cases}
\usepackage[utf8]{inputenc}
\usepackage{url}
\usepackage{longtable}
\usepackage{xspace}
\usepackage{acronym}
\usepackage{aas_macros}

\usepackage{amsfonts}
\usepackage{xfrac}
\usepackage{pifont}
\usepackage{fourier}
\usepackage{hyperref}
\usepackage{bm}
\usepackage{enumitem}

\definecolor{rossoferrari}{HTML}{D9073D}
\definecolor{mediumblue}{HTML}{0000CD}
\definecolor{forestgreen}{HTML}{228B22}
\definecolor{desy_blue}{HTML}{009EE2}
\definecolor{desy_orange}{HTML}{FD8800}
\definecolor{light_pink}{rgb}{1,0.4,0.4}
\definecolor{light_blue}{rgb}{0.284602,0.317763,0.963947}
\hypersetup{setpagesize=false,bookmarksnumbered=true,bookmarksopen=true,%
colorlinks=true,linkcolor=light_blue,urlcolor=rossoferrari,citecolor=rossoferrari,linktocpage=false}

\newcommand{\Mpl}{M_\mathrm{Pl}}
\newcommand{\ns}{n_{\mathrm{s}}}

\newcommand{\uc}{\mathrm{c}}

\newcommand{\calN}{\mathcal{N}}

\newcommand{\calP}{\mathcal{P}}

\newcommand{\Ne}{\mathcal{N}}

\newcommand{\HI}{H_{\rm inf}}
\newcommand{\g}{\ {\rm g}}
\newcommand{\Mpc}{\ {\rm Mpc}}

\newcommand{\beae}[1]{\begin{equation}\begin{aligned} #1 \end{aligned}\end{equation}}

\newcommand{\bae}[1]{\begin{align} #1 \end{align}}

\newcommand{\bmbe}[1]{\begin{multlined}[b] #1 \end{multlined}}

\definecolor{MONZA}{HTML}{CF000F}
\definecolor{DARKBLUE}{HTML}{00008b}
\definecolor{DARKMAGENTA}{HTML}{8b008b}

\def\beqa{\begin{eqnarray}}
\def\eeqa{\end{eqnarray}}

\def\del{\partial}
\newcommand{\GeV}{\ {\rm GeV}}

\def\lmk{\left(}
\def\rmk{\right)}
\def\lkk{\left[}
\def\rkk{\right]}
\def\la{\left<}
\def\ra{\right>}
\newcommand{\eq}[1]{Eq.~(\ref{#1})}
\def\beq#1\eeq{\begin{align}#1\end{align}}

\newcommand{\noise}{\xi}
\newcommand{\infl}{\zeta}
\newcommand{\wt}{\chi}
\newcommand{\curv}{\mathcal{R}}

\begin{document}

\preprint{TU-1184}

\title{On the primordial black hole formation in hybrid inflation}

\author{Yuichiro Tada}
\email{tada.yuichiro.y8@f.mail.nagoya-u.ac.jp}
\affiliation{Institute for Advanced Research, Nagoya University,
Furo-cho Chikusa-ku, 
Nagoya 464-8601, Japan}
\affiliation{Department of Physics, Nagoya University, 
Furo-cho Chikusa-ku,
Nagoya 464-8602, Japan}
\affiliation{Theory Center, IPNS, KEK, 
1-1 Oho, Tsukuba, 
Ibaraki 305-0801, Japan}

\author{Masaki~Yamada}
\email{m.yamada@tohoku.ac.jp}
\affiliation{Department of Physics, Tohoku University, Sendai, Miyagi 980-8578, Japan}
\affiliation{FRIS, Tohoku University, Sendai, Miyagi 980-8578, Japan}

\date{\today}

\begin{abstract}
\noindent
We revisit the scenario of primordial black hole (PBH) formation from large curvature perturbations generated during the waterfall phase transition in hybrid inflation models. In a minimal setup considered in the literature, the mass and abundance of PBHs are correlated and astrophysical size PBHs tend to be overproduced. This is because a longer length scale for curvature perturbations (or a larger PBH mass) requires a longer waterfall regime with a flatter potential, which results in overproduction of curvature perturbations. However, in this paper, we discuss that the higher-dimensional terms for the inflaton potential affect the dynamics during the waterfall phase transition and show that astrophysical size PHBs of the order of $10^{17\text{--}23} \, {\rm g}$ (which can explain the whole dark matter) can form in some parameter space consistently with any existing constraints. The scenario can be tested by observing the induced gravitational waves from scalar perturbations by future gravitational wave experiments, such as LISA. 
\end{abstract}\acresetall

\maketitle


\acrodef{CMB}{cosmic microwave background}
\acrodef{BH}{black hole}
\acrodef{PBH}{primordial black hole}
\acrodef{DM}{dark matter}
\acrodef{GW}{gravitational wave}

\section{Introduction}

The seeds of the structure of the Universe can be generated by quantum fluctuations of inflatons or curvatons during inflation.
The amplitude of curvature perturbations is of the order of $10^{-5}$ at the \ac{CMB} scale~\cite{Planck:2018vyg}, whereas larger curvature perturbations may be generated at a smaller scale~\cite{Ivanov:1994pa, GarciaBellido:1996qt, Kawasaki:1997ju, Yokoyama:1998pt, Garcia-Bellido:2017mdw, Hertzberg:2017dkh}. 
In fact, observations of supermassive \ac{BH}~\cite{LyndenBell:1969yx, Kormendy:1995er} and \ac{BH} merger events by gravitational wave detectors~\cite{LIGOScientific:2016dsl, LIGOScientific:2021djp} imply the existence of \acp{PBH} that are generated via the collapse of overdense regions~\cite{Hawking:1971ei,Carr:1974nx,Carr:1975qj}.%
\footnote{
There are some other scenarios to have a PBH formed: cosmic strings~\cite{Hawking:1987bn, Garriga:1992nm, Caldwell:1995fu}, bubble collisions~\cite{Hawking:1982ga}, domain walls~\cite{Garriga:1992nm, Khlopov:2008qy, Garriga:2015fdk, Deng:2016vzb}, and collapse of vacuum bubbles~\cite{Garriga:2015fdk, Deng:2017uwc, Deng:2018cxb,Deng:2020mds}. }
The \ac{PBH} is also a candidate for \ac{DM}~\cite{Chapline:1975ojl} if its mass is within $10^{17\text{--}23} \g$ (see, e.g., Refs.~\cite{Carr:2016drx,Inomata:2017okj,Inomata:2017vxo}). 
Such large curvature perturbations can be generated if the inflaton or a spectator field goes through a very flat potential during inflation (see also Ref.~\cite{Escriva:2022duf} for a recent review of \ac{PBH}s).

One of the simplest examples to generate large curvature perturbations is the hybrid inflation model, where inflation ends by a waterfall (second-order) phase transition~\cite{Linde:1993cn}.%
\footnote{
See Refs.~\cite{GarciaBellido:1996qt,Kawasaki:1997ju,Yokoyama:1998pt,Kawasaki:2006zv,Kawaguchi:2007fz,Kohri:2007qn,Frampton:2010sw,Drees:2011yz,
Kawasaki:2012kn,Kawasaki:2012wr,Kawasaki:2016pql,Inomata:2016rbd,Ezquiaga:2017fvi,Kannike:2017bxn,Germani:2017bcs,Motohashi:2017kbs,Ballesteros:2017fsr,Hertzberg:2017dkh,Cicoli:2018asa,Cheong:2019vzl,Ballesteros:2020qam,Pi:2021dft,Geller:2022nkr} for other models. 
}
The waterfall field can have a flat potential to generate large curvature perturbations~\cite{Clesse:2010iz,Kodama:2011vs,Mulryne:2011ni}. 
Since the waterfall phase transition happens at the last stage of inflation, 
this results in large curvature perturbations at relatively small scales. 
Although one can make its scale larger by flattening the potential of the waterfall field, 
the amplitude of curvature perturbations then becomes too large. 
\ac{PBH} mass and abundance are correlated in the minimal setup, which results in an overproduction of astrophysical size \acp{PBH}. 
This naive picture is actually confirmed by analytical calculation and numerical calculations in the stochastic formalism~\cite{Clesse:2015wea,Kawasaki:2015ppx}.

In this paper,
we point out that 
\acp{PBH} with an astrophysical size can be generated in a simple hybrid inflation model, by demonstrating that the quadratic and cubic terms for the inflaton potential affect curvature perturbations, which are omitted in the literature. 
In particular, 
the degeneracy between the \ac{PBH} mass and its abundance can be removed by those effects and the peak amplitude of curvature perturbations can be reduced by tuning parameters. 
We discuss how much tuning is required to predict the desired amount of PBHs.
Moreover, the spectral index of curvature perturbations at the \ac{CMB} scale can be consistent with observations.

The organization of this paper is as follows. 
In Sec.~\ref{sec:model}, 
we briefly review the analytic calculation for curvature perturbations, following Ref.~\cite{Clesse:2015wea} 
and clarify that the resulting spectrum has degeneracy between its peak amplitude and corresponding wavenumber if we omit quadratic and cubic terms for the inflaton potential. 
In Sec.~\ref{sec:analytic}, we take into account quadratic and cubic terms for the inflaton potential and show that 
the degeneracy can be resolved by those effects. 
In Sec.~\ref{sec:numerical}, we solve the classical equation of motion numerically and search parameter space that predicts a desired amplitude of curvature perturbations. 
We then consider \ac{PBH} formation in Sec.~\ref{sec:PBH} and show that \acp{PBH} with mass $10^{17\text{--}23}\g$ can be generated consistently with all existing constraints. 
We also show that observable \acp{GW} are generated from the second-order curvature perturbations. 
Sec.~\ref{sec:conclusions} is devoted to discussion and conclusions.



\section{Hybrid inflation model}
\label{sec:model}

We consider the hybrid inflation model~\cite{Clesse:2010iz,Kodama:2011vs,Mulryne:2011ni,Clesse:2015wea,Kawasaki:2015ppx}
\beq
    V(\phi,\psi)
    =\Lambda^4
    \lkk
    \pqty{1-\frac{\psi^2}{M^2}}^2
    +2\frac{\phi^2\psi^2}{\phi_\uc^2M^2}
    +V(\phi)
    \rkk \,,
    \label{eq:potential}
\eeq
where $\phi$ is an inflaton and $\psi$ is a waterfall field. 
We are interested in the dynamics of fields around the waterfall phase transition, where curvature perturbations are generated for the scales of interest. 
The inflaton potential $V(\phi)$ is expanded around the critical point $\phi_\uc$ as 
\beq
 V(\phi) = \frac{\phi-\phi_\uc}{\mu_1}
    -\frac{(\phi-\phi_\uc)^2}{\mu_2^2}
    + \frac{(\phi-\phi_\uc)^3}{\mu_3^3} \,,
\eeq
where 
$\Lambda$, $M$, $\phi_\uc$, $\mu_1$, $\mu_2$, and $\mu_3$ are dimensionful parameters. 
The curvature along with the waterfall direction changes its sign when the inflaton $\phi$ reaches the critical point. We denote the time at which $\phi = \phi_\uc$ as the waterfall phase transition. 
The Hubble parameter during inflation is $\HI^2 \simeq \Lambda^4 / (3 \Mpl^2)$. 
We extend the model in Refs.~\cite{Clesse:2015wea,Kawasaki:2015ppx} by introducing the cubic potential in $V(\phi)$. 
We will see that the cubic term plays an important role to obtain a desired amplitude of curvature perturbations as well as the observed spectral index. 

In this section, we review the calculation for curvature perturbations generated from waterfall fields by an analytic method used in Ref.~\cite{Clesse:2015wea}, omitting quadratic and cubic terms in the inflaton potential. 
In the next section, we include the effect of those terms.

\subsection{Spectrum at the CMB scale}

We first analyse the dynamics of inflaton before the waterfall phase transition, where $\phi > \phi_\uc$. 
In this regime, we can solve the equation of motion for $\phi$ by approximating $\psi \simeq 0$.

We want to calculate the spectral index and amplitude of curvature perturbations at the \ac{CMB} scale. 
We denote the backward e-folding number at the \ac{CMB} scale and the one at the waterfall phase transition as $\Ne_*$ and $\Ne_\uc$, respectively. 
The observed amplitude of curvature perturbations at the pivot scale 
is given by 
\beq
 {\cal P}_\curv (k_*) \simeq 2.1 \times 10^{-9}  \,,
 \label{COBE}
\eeq
where $k_*$ ($= 0.05 \, {\rm Mpc}^{-1}$) represents the wavenumber at the pivot scale~\cite{Planck:2018vyg}. 
These perturbations exit the horizon 
before the waterfall phase transition, 
and should come from 
the fluctuation of the inflaton $\phi$. 
Its amplitude is calculated from 
\beq 
 {\cal P}_\curv = \frac{\HI^2}{8 \pi^2 \epsilon \Mpl^2} \,,
\eeq
where 
\beq 
 \epsilon = \frac12 \lmk \Mpl \frac{V_\phi}{V} \rmk^2 
 \simeq \frac{\Mpl^2}{2 \mu_1^2} \,. 
\eeq
From \eq{COBE}, 
we require 
\beq
 \Lambda \simeq 1.7 \times 10^{14} \GeV 
 \lmk \frac{\mu_1}{10^5 \Mpl} \rmk^{-1/2}  \,,
\eeq
or 
\beq
 \HI \simeq 
 7.0 \times 10^{9} \GeV 
 \lmk \frac{\mu_1}{10^5 \Mpl} \rmk^{-1} \,. 
\eeq
The e-folding number at the pivot scale is given by 
\beq
 \Ne_*  &\simeq 51.5
 + \frac{1}{2} \ln \lmk \frac{\HI}{7 \times 10^{9} \GeV} \rmk 
 + \frac{1}{6} \ln \lmk \frac{H_{\rm RH}}{\HI} \rmk \,, 
 \label{eq:Nstar}
\eeq
where $H_{\rm RH}$ represents the Hubble parameter at the completion of reheating.

The spectral index is given by 
\beq
 \ns &= 1 + \eval{2\Mpl^2 \frac{V_{\phi \phi} }{V}}_{k_*} 
 \simeq 1 - 4  \frac{\Mpl^2}{\mu_2^2}  \,, 
\eeq
where we neglected the contribution of $\epsilon$ compared to that of $V_{\phi\phi}$ and $\phi_*$ represents the field value of $\phi$ at $\Ne = \Ne_*$.

\subsection{Stochastic effect around the waterfall phase transition}

Around the critical point $\phi = \phi_\uc$, the curvature of the potential along with $\psi$ direction is so small that its quantum fluctuations efficiently grow with time. 
It obeys the slow-roll Langevin equation (see Refs.~\cite{Starobinsky:1982ee,Starobinsky:1986fx,Nambu:1987ef,Nambu:1988je,Kandrup:1988sc,Nakao:1988yi,Nambu:1989uf,Mollerach:1990zf,Linde:1993xx,Starobinsky:1994bd} for the first papers on the subject) 
\bae{
    \partial_N \psi =-\Mpl^2\frac{V_\psi}{V}+\frac{1}{2\pi}\sqrt{\frac{V}{3\Mpl^2}}\noise(N)  \,,
    \label{eq:stochastic_r}
}
where $N$ is the forward e-folding number as the time variable and $\noise$ is the independent noise $\braket{\noise(N)\noise(N^\prime)}=\delta(N-N^\prime)$. 
We can neglect the noise term for the inflaton $\phi$ around the waterfall phase transition for our purpose.

The noise term for $\psi$ is important at a time around and before the waterfall phase transition but can be negligible at a later time. 
We divide the dynamical regime into two phases: the stochastic phase and classical phase~\cite{Clesse:2015wea}. Initially, the noise term dominates $\psi$'s dynamics.%
\footnote{
This regime was omitted in Ref.~\cite{Spanos:2021hpk}, where they set the initial condition for the classical phase by hand. This is the reason a desirable mass of \acp{PBH} was obtained in the hybrid inflation model, even if the cubic  and higher-order terms in the inflaton potential are irrelevant for the dynamics. However, this is not allowed if one correctly considers the stochastic dynamics. 
In fact, the probability that $\la \psi^2 \ra$ is deviated from \eq{eq:r0} must be exponentially suppressed because of the following reason. 
The relevant mode exits the horizon at the e-folding number of $\Ne_\uc$ ($\sim 10$). The number of Hubble-volume patches corresponding to that mode within the present observable Universe is then of the order of $e^{3(\Ne-\Ne_\uc)}$, where $\Ne \sim 60$ is the total e-folding number for the observable Universe. The value of $\la \psi^2 \ra$ is calculated from the ensemble average over those patches, so that the probability for deviation from its averaged value is exponentially suppressed by a factor of $e^{-(3/2)(\Ne-\Ne_\uc)}$. 
Therefore one should not take a different value of $\psi$ at the waterfall phase transition from $\psi_{0}$ by hand. 
If one used a different (wrong) value, the resulting relation between $\mathcal{P}_\curv$ and $\Ne_\uc$, which we will see shortly, would be modified accordingly. 
We then conclude that the result is extremely unrealistic if the field value of $\psi$ at the waterfall phase transition is different from $\psi_0$. 
\label{footnote2}
}
This sets the initial condition for the classical phase, where the classical equation of motion (i.e., the first term in the right-hand side in \eq{eq:stochastic_r}) dominates the dynamics. 
We solve the dynamics of $\psi$ and $\phi$, and calculate the curvature perturbations by the $\delta \Ne$ formalism~\cite{Starobinsky:1985ibc,Salopek:1990jq,Sasaki:1995aw,Sasaki:1998ug,Lyth:2004gb}.

We denote the forward e-folding number 
at the time of the waterfall phase transition (at which $\phi  = \phi_\uc$) as $N_\uc$. 
Let us consider the dynamics around $N \approx N_\uc$. 
If we neglect the stochastic noise for $\phi$, its solution is given by 
\beq
 \phi \simeq \phi_\uc - \frac{\Mpl^2 (N-N_\uc)}{ \mu_1}  \,,
\eeq
for $N \approx N_\uc$. 
From \eq{eq:stochastic_r}, 
the equation of motion for $\la \psi^2 \ra$ is given by 
\beq
 \dv{N} \la \psi^2 \ra = \lmk \frac{4}{\Pi} \rmk^2 (N-N_\uc) \la \psi^2 \ra +  \frac{\HI^2}{4 \pi^2}  \,,
\eeq
where we define 
\beq
 \Pi \equiv \frac{M \sqrt{\mu_1 \phi_\uc}}{\Mpl^2} \,. 
 \label{Pi}
\eeq
The first term in the right-hand side represents the classical force, 
while the second term represents the stochastic force from the noise term. 
The solution to this equation can be written by the error function such as 
\beq
\label{eq:r}
 \la \psi^2 \ra (N) = 
 \psi_{0}^2 \lkk 1 + {\rm Erf}\lmk \frac{2\sqrt{2} (N-N_\uc)}{\Pi} \rmk \rkk \exp \lkk \frac{8 (N-N_\uc)^2}{\Pi^2} \rkk  \,,
\eeq
where 
${\rm Erf}(x) \equiv (2/\sqrt{\pi}) \int_0^x e^{-t^2} \dd{t}$ 
and 
we use $\la \psi^2 \ra \approx 0$ for $N \to -\infty$. 
The amplitude of $\la \psi^2 \ra$ at the waterfall phase transition ($N=N_\uc$) is given by 
\beq
 \psi_{0}^2 = \frac{ \Lambda^4 \Pi}{48 \sqrt{2 \pi^3} \Mpl^2 }  \,,
 \label{eq:r0}
\eeq
where we adopted $\HI^2 \simeq \Lambda^4 / 3 \Mpl^2$.

The value of \eq{eq:r0} can be used as the initial condition for the field $\psi$ at the waterfall phase transition. 
Then we can calculate the spectrum of curvature perturbations by solving the classical equations of motion for $N > N_\uc$ with the ``initial'' condition of $\psi(N_\uc) = \psi_0$ and $\phi(N_\uc) = \phi_\uc$.

\subsection{Dynamics after the waterfall phase transition}

Following Refs.~\cite{Kodama:2011vs,Clesse:2015wea}, we analytically consider the dynamics after the waterfall phase transition. 
We omit the second term in the equation of motion for $\psi$ in \eq{eq:stochastic_r} and solve the classical equation of motion for $N >N_\uc$. 
We denote $\la \psi^2 \ra$ as $\psi^2$ for notational simplicity. 
The initial conditions are given by $\phi = \phi_\uc$ and $\psi = \psi_{0}$, with $\psi_{0}$ given by \eq{eq:r0}.

We introduce the following notations: 
\beq
 \phi \equiv \phi_\uc e^\infl \simeq \phi_\uc ( 1 + \infl), \quad
 \psi \equiv \psi_{0} e^{\wt} \,. 
\eeq
Here we assume $\infl \ll 1$. 
This is actually the case until the end of inflation, where 
\beq
 \abs{ \frac{V_{\psi \psi} \Mpl^2}{V} } = c_{\rm EoI} = \mathcal{O}(1)
 \quad \leftrightarrow \quad 
 - \infl_{\rm end} \simeq \frac{c_{\rm EoI} M^2}{8 \Mpl^2} \ll 1 \,, 
 \label{xi_end}
\eeq
with an $\mathcal{O}(1)$ parameter $c_{\rm EoI}$ that determines the end of inflation.

The equations of motion for $\infl$ and $\wt$ are expressed, respectively, as 
\beq
\label{chiEOM1}
& \dv{N} \infl = - \frac{ \Mpl^2}{\mu_1 \phi_\uc} 
 - \frac{4 \Mpl^2}{\phi_\uc^2 M^2} \psi_{0}^2 e^{2 \wt} \,, 
 \\
& \dv{N} \wt = - \frac{8 \Mpl^2}{M^2} \infl \,, 
\label{xiEOM1}
\eeq
where we use the slow-roll approximation to neglect the second derivatives.

We note that the last term in \eq{chiEOM1} 
is negligible initially 
and then dominates at a later epoch. 
We thus decompose the waterfall in two phases, 
such that the first term dominates in the first phase 
and the last term dominates in the second phase. 
The threshold between the two phases, denoted by a subscript 2, is determined by 
\beq
\frac{\Mpl^2}{\mu_1 \phi_\uc} = \frac{4 \Mpl^2}{\phi_\uc^2 M^2} \psi_{0}^2 e^{2 \wt_2}
 \quad \leftrightarrow \quad
 \wt_2 = \ln \lmk \frac{M \sqrt{\phi_\uc}}{2 \psi_{0} \sqrt{\mu_1}} \rmk \,. 
 \label{chi2}
\eeq
Using Eqs.~(\ref{COBE}) and (\ref{chi2}), 
we obtain 
\beq
 \wt_2 
 \simeq 
 \frac12 \ln \lmk     \sqrt{\frac{2}{\pi}}\frac{\Pi}{2.1 \times 10^{-9}} \rmk
 \simeq 
 9.9 + \frac12 \ln \Pi \,. 
\eeq
Thus we expect $\wt_2 \sim 10$.

In the first phase, $\wt \ll \wt_2$ 
and the last term in \eq{chiEOM1} is negligible. 
The solution to the coupled equations is given by 
\beq
 &\infl (N) \simeq - \frac{\Mpl^2}{\mu_1 \phi_\uc} (N-N_\uc) \,, 
 \\
 &\wt (N) \simeq  \frac{4 \Mpl^2}{M^2} \frac{\Mpl^2}{\mu_1 \phi_\uc} (N-N_\uc)^2  \,. 
 \label{chi-1}
\eeq
Denoting the e-folding number at the end of the first phase as $N_1$, 
we obtain the e-folding number during the first phase such as 
\beq
 N_1- N_\uc &\simeq 
 \frac{M \sqrt{\mu_1 \phi_\uc}}{2 \Mpl^2} \wt_2^{1/2}  \,,
\eeq
where we adopt \eq{chi-1}. 
The value of $\infl$ at the end of the first phase is given by%
\footnote{
We note that $N_1$ represents $N$ at the end of the first phase, 
whereas $\wt_2$ and $\infl_2$ represent $\wt$ and $\infl$, respectively, at the beginning of the second phase. 
Since the end of first phase is identical to the beginning of the second phase, 
$\infl_2 = \infl(N_1)$. 
This would be confusing, but we adopt this notation following Ref.~\cite{Clesse:2015wea}. 
}
\beq
 \infl_2 \simeq - \frac{M}{2 \sqrt{\mu_1 \phi_\uc}} \wt_2^{1/2}  \,.
 \label{xi2}
\eeq

In the second phase, $\wt \gg \wt_2$ 
and the first term is negligible in \eq{chiEOM1}. 
The solution is given by 
\beq
 \infl^2(N) = \infl_2^2 + \frac{M^2}{8 \mu_1 \phi_\uc} \lkk e^{2(\wt(N)-\wt_2)} - 1 \rkk \,. 
 \label{xi-2}
\eeq
We expect that 
the solutions in the two phases are simply connected at $\wt = \wt_2$ 
with $\infl = \infl_2$. 
Denoting the e-folding number at the end of the second phase (i.e., at the end of inflation) as $N_2$, 
we obtain the e-folding number during the second phase such as 
\beq
 N_2 - N_1 &= 
 - \int_{\wt_2}^{\wt_{\rm end}} \frac{M^2}{8 \Mpl^2 \infl} \dd{\wt}
 = \frac{M^2}{8\Mpl^2} \frac{c}{\abs{\infl_2}} \,, 
 \label{eq:N2}
\eeq
where we use Eqs.~(\ref{chiEOM1}) and (\ref{xi-2}), 
and define 
\beq
\label{eq:c}
 c \equiv \int_{\wt_2}^{\wt_{\rm end}} \frac{\dd{\wt}}{\sqrt{1 + M^2 / (8 \mu_1 \phi_\uc \infl_2^2) \lkk e^{2 (\wt - \wt_2)} - 1 \rkk}} \,. 
\eeq
One can demonstrate that $c < 1$ and it is asymptotic to unity in the limit of $\wt_{\rm end} \gg \wt_2$.

In summary, 
the total e-folding number from the waterfall phase transition to the end of inflation, $\Ne_\uc$, is given by 
\beq
 \Ne_\uc &= N_2 - N_\uc
 =
 \Pi \lmk \frac{\sqrt{\wt_2}}{2} + \frac{c}{4 \sqrt{\wt_2}} \rmk \,, 
\eeq
where we use Eqs.~(\ref{xi2}) and (\ref{Pi}).

\subsection{Spectrum at a small scale}

Now we can calculate the curvature perturbations from 
the fluctuation of the waterfall fields after the waterfall phase transition. 
According to the $\delta \Ne$ formalism~\cite{Starobinsky:1985ibc,Salopek:1990jq,Sasaki:1995aw,Sasaki:1998ug,Lyth:2004gb}, 
the power spectrum of the curvature perturbation $\curv$ can be calculated from 
\beq
 {\cal P}_\curv 
 &= \lmk \frac{\del \Ne_k}{\del \psi_{k}}\rmk^2 \lmk \frac{\HI}{2\pi} \rmk^2 \,, 
 \label{eq:deltaN}
\eeq
where $\psi_{k}$ represents the value of $\psi$ at which the mode with a wavenumber $k$ exits the horizon at $\Ne = \Ne_k$. 
Here, $\Ne_k$ is understood as $\Ne_k (\phi_k, \psi_k)$ with $\phi_k$ and $\psi_k$ following the classical trajectory. The derivative $\del \Ne_k / \del \psi_k$ means the partial derivative with respect to $\psi_k$ with a fixed $\phi_k$. 

Within the first phase, we note that 
\beq
 \wt \simeq \frac{4}{\Pi^2} (N-N_k)^2 - \frac{8\Mpl^2 }{M^2} \infl_k (N-N_k)  + \wt_k \,, 
\eeq
and hence 
\beq
 \Ne_k \equiv N_2 - N_k \simeq 
 \frac{\Pi}{2} \sqrt{\chi_2 - \chi_k + 8 \zeta_k (N_2-N_k) \Mpl^2 / M^2} \,,
\eeq
where we neglected the e-folding from the second phase. 
Taking the partial derivative with respect to $\wt_k$, 
we obtain
\beq
\label{delNdelphi}
 \frac{\del \Ne_k(\wt_k)}{\del \psi_{k}} \simeq  
 - \frac{\Pi}{4\, \psi_{0} e^{\wt_k}} 
 \frac{1}{\sqrt{\wt_2}}  \,. 
\eeq
Here we neglect the term with $\infl_k$ because the dominant $\chi_k$ dependence comes from the exponential factor. 
Here, $\chi_k$ can be rewritten in terms of e-folding number such as 
\beq 
 \wt_k &\simeq \frac{4}{\Pi^2} ( \Ne_\uc - \Ne_k )^2 \simeq \frac{\wt_2}{\Ne_1^2} ( \Ne_\uc - \Ne_k )^2 \,. 
  \label{N12}
\eeq
We note that this procedure is justified only for $\wt_k \ge 0$. 
The maximal value is obtained at this threshold such as 
\beq
 {\cal P}_\curv^{(\rm peak)}
 &\simeq 
 \frac{\Pi}{2 \sqrt{2\pi} \wt_2 } 
\simeq 0.013 \, \Ne_\uc \lmk \frac{\wt_2}{10} \rmk^{-3/2} \,, 
 \label{Pzeta}
\eeq
where we use $\wt_2 \gg 1$ and neglect $c/\wt_2 \ll 1$ in the second line. 
The shape of the power spectrum is determined by $e^{\wt_k}$ in \eq{delNdelphi} such as 
\beq
{\cal P}_\curv (\Ne_k) 
&\simeq {\cal P}_\curv^{(\rm peak)} e^{- 2 \wt_k} \,, 
\label{width}
\eeq
where 
$\wt_k$ can be written in terms of $\Ne_k (\wt_k)$ by using \eq{N12}. 
The leading term is the Gaussian form with the width of $\Ne_1 / (2\sqrt{\wt_2})$ ($\sim \Ne_1 / 6$).%
\footnote{
We implicitly assume that inflation ends during phase 2, namely, $\abs{\infl_{\rm end}} > \abs{\infl_2}$. If one adopts the condition of $c_{\rm EoI} = 1$ in \eq{xi_end} for the end of inflation, the condition is not satisfied for $\Pi^2 \lesssim 16 \wt_2 \simeq 160$. 
However, we note that inflation continues 
for $\Delta \Ne \sim 1$ and $\abs{\infl}$ grows much even after $\abs{\eta}$ becomes as large as unity. 
Moreover, the ambiguity for the end of inflation 
does not affect our results because 
it comes into a factor of $c$ in \eq{eq:c} 
and its dependence is negligible in \eq{Pzeta}. 
We thus need to take $c_{\rm EoI}$ a little bit larger than unity to calculate the spectrum of curvature perturbations.
}

From Eqs.~(\ref{Pzeta}) and (\ref{width}), 
we observe that the peak amplitude of curvature perturbations ${\cal P}_\curv^{(\rm peak)}$ 
and its wavenumber 
are determined by the e-folding number from the end of inflation $\Ne_\uc$. 
Therefore the peak amplitude and corresponding peak wavelength are related with each other in the leading order calculation. 
Thus we cannot obtain a desirable amount of curvature perturbations to generate PBHs with astrophysical scales in the minimal model, where the second and third terms in the last parenthesis in \eq{Pzeta} are neglected~\cite{Clesse:2015wea}. 
This No-Go theorem for massive \acp{PBH} in the mild-waterfall hybrid inflation is also confirmed in the full-numerical stochastic-$\delta\calN$ approach~\cite{Kawasaki:2015ppx}.


\section{Effect of quadratic and cubic terms}
\label{sec:analytic}

We now extend the analysis to include the next-to-leading order effect from the quadratic and cubic terms of the inflaton potential, which becomes important for a large $\Pi$. 

\subsection{Spectral index}

The spectral index is given by 
\beq
 \ns 
 \simeq 1 - 4  \frac{\Mpl^2}{\mu_2^2} 
 + 12 \frac{\Mpl^2 \lmk \phi_* - \phi_\uc \rmk}{\mu_3^3} \,,
 \label{ns}
\eeq
including the cubic term.
We will see shortly that either $\mu_2$ or $\mu_3$ is determined in order to suppress the peak amplitude of curvature perturbations at a smaller scale. 
We can then take the other parameter appropriately to make $\ns$ within the observed value $\ns = 0.9649 \pm 0.0042$~\cite{Planck:2018vyg}. 
We also calculate 
the running of spectral index $\dd{\ns}/\dd{\ln k}$ 
and check that $0 > \dd{\ns}/\dd{\ln k} \gtrsim -0.0026$ in our parameter of interest. 
This is consistent with the current constraint of $\dd{\ns}/\dd{\ln k} = -0.0045 \pm 0.0067$~\cite{Planck:2018vyg}.

We can determine $\phi_*$ by inversely solving its  equation of motion from $\Ne = \Ne_\uc$
to $\Ne = \Ne_*$. 
This is justified under the slow-roll approximation, where the initial condition for the velocity is irrelevant for the dynamics. 
In Sec.~\ref{sec:numerical}, we numerically calculate $\phi_*$ and $\ns$ for each parameter set after calculating $\Ne_\uc$. 

\subsection{Dynamics after the waterfall phase transition}

The equations of motion for $\infl$ and $\wt$ are now expressed as 
\beq
\label{chiEOM}
& \dv{N} \infl = - \frac{\Mpl^2}{\mu_1 \phi_\uc} + \frac{2 \Mpl^2}{\mu_2^2} \infl
- \frac{3 \phi_\uc \Mpl^2}{\mu_3^3} \infl^2
 - \frac{4 \Mpl^2}{\phi_\uc^2 M^2} \psi_{0}^2 e^{2 \wt} \,, 
 \\
& \dv{N} \wt = - \frac{8 \Mpl^2}{M^2} \infl \,, 
\label{xiEOM}
\eeq
under the slow-roll approximation. 
We include the second and third terms in the right-hand side in \eq{chiEOM} as perturbations under the following approximation: 
\beq
 \frac{\Mpl^2 (N-N_\uc) }{ \mu_2^2} \ll 1, \quad
 \frac{\Mpl^4 (N-N_\uc)^2}{\mu_1 \mu_3^3} \ll 1 \,. 
\eeq 
We consider their effects up to next-to-leading terms.

We first note that the corrections to $\wt_2$ from $1/\mu_2^2$ and $1/\mu_3^3$ terms are only logarithmic and are negligible. 
In the first phase, $\wt \ll \wt_2$, the solution to the coupled equations is given by 
\beq
 &\bmbe{\infl (N) \simeq - \frac{\Mpl^2}{\mu_1 \phi_\uc} (N-N_\uc)
 - \frac{\Mpl^2}{\mu_2^2} \frac{\Mpl^2}{\mu_1 \phi_\uc} (N-N_\uc)^2
 \\
 - \frac{\Mpl^2 \phi_\uc}{\mu_3^3} \lmk \frac{\Mpl^2}{\mu_1 \phi_\uc} \rmk^2 (N-N_\uc)^3 \,,} 
 \\
 &\bmbe{\wt (N) \simeq  \frac{4 \Mpl^2}{M^2} \frac{\Mpl^2}{\mu_1 \phi_\uc} (N-N_\uc)^2 
 +\frac{8 \Mpl^4}{3 M^2 \mu_2^2} \frac{\Mpl^2}{\mu_1 \phi_\uc} (N-N_\uc)^3 
 \\
 + \frac{2 \Mpl^4 \phi_\uc}{ M^2 \mu_3^3} \lmk \frac{\Mpl^2}{\mu_1 \phi_\uc} \rmk^2 (N-N_\uc)^4
  \,.}
 \label{chi-1-2}
\eeq
The e-folding number during the first phase is found as
\beq
 N_1 - N_\uc \simeq 
 \frac{M \sqrt{\mu_1 \phi_\uc}}{2 \Mpl^2} \wt_2^{1/2} 
 - \frac{\mu_1 \phi_\uc M^2}{12 \mu_2^2 \Mpl^2} \wt_2 
 - \frac{\phi_\uc M^3 \sqrt{\mu_1 \phi_\uc} }{32 \mu_3^3 \Mpl^2} \wt_2^{3/2} \,, 
\eeq
where we adopt \eq{chi-1-2}. 
The value of $\infl$ at the end of the first phase is given by 
\beq
 \infl_2 \simeq - \frac{M}{2 \sqrt{\mu_1 \phi_\uc}} \wt_2^{1/2} - 
 \frac{M^2}{6 \mu_2^2} \wt_2 
 - \frac{3 \phi_\uc M^3}{ 32\mu_3^3 \sqrt{\mu_1 \phi_\uc}} \wt_2^{3/2} \,. 
 \label{xi2-2}
\eeq
The dynamics in the second phase does not change qualitatively. 
The solution and the e-folding number  are again given by Eqs.~(\ref{xi-2}) and (\ref{eq:N2}).

In summary, 
the total e-folding number from the waterfall phase transition to the end of inflation, $\Ne_\uc$, is given by 
\beq
 \Ne_\uc &= N_2 - N_\uc
 \nonumber \\
 &\bmbe{=
 \Pi \lmk \frac{\sqrt{\wt_2}}{2} + \frac{c}{4 \sqrt{\wt_2}} \rmk 
 - \Pi^2 \frac{\Mpl^2}{12 \mu_2^2} \lmk \wt_2 + c \rmk
 \\
 - \Pi^3 \frac{\Mpl^4}{32 \mu_1 \mu_3^3 } \lmk \wt_2^{3/2} + \frac{3 c}{2} \wt_2^{1/2} \rmk \,,} 
 \label{eq:NPT}
\eeq
where we use Eqs.~(\ref{xi2-2}) and (\ref{Pi}).

\subsection{Spectrum of curvature perturbations}

Now we calculate the curvature perturbations from 
the fluctuation of the waterfall fields after the waterfall phase transition. 
First, $\wt_k$ is solved as 
\beq 
 \wt_k &= \frac{4}{\Pi^2} \lkk ( \Ne_\uc - \Ne_k )^2 + \frac{2 \Mpl^2}{3 \mu_2^2} ( \Ne_\uc - \Ne_k )^3 
 \right.
 \nonumber \\
  &\qquad \qquad \qquad \qquad \qquad \left. + \frac{\Mpl^4}{2 \mu_1 \mu_3^3} ( \Ne_\uc - \Ne_k )^4
  \rkk  \,,
  \label{N122}
\eeq
within the first phase.
For $\chi > \chi_k$, $\chi(N)$ can be given by 
\beq 
 \wt - \wt_k &\simeq \frac{4}{\Pi^2} \lkk ( N - N_k )^2 + \frac{2 \Mpl^2}{3 \mu_2^2} ( N - N_k )^3 
 \right.
 \nonumber \\
  &\qquad \qquad \qquad  \left. + \frac{\Mpl^4}{2 \mu_1 \mu_3^3} (N - N_k )^4 + \mathcal{O}(\zeta_k)
  \rkk  \,, 
  \label{N1222}
\eeq
where we neglect the term with $\zeta_k$ for simplicity. We calculate the peak amplitude of curvature perturbations 
because, as we discussed around \eq{delNdelphi}, the dominant $\chi_k$ dependence comes from the exponential factor $e^{-\chi_k}$. 
Noting that $\Ne_k \equiv N_2 - N_k$,
we obtain the partial derivative of $\Ne_k$ with respect to $\psi_k$ such as 
\beq
 \frac{\del \Ne_k(\wt_k)}{\del \psi_{k}} \simeq  
 - \frac{\Pi}{4\, \psi_{0} e^{\wt_k}} 
  \lmk 
 \frac{1}{\sqrt{\wt_2}} -   \frac{\Mpl^2}{3 \mu_2^2} \Pi
 - \frac{3 \Mpl^4}{16 \mu_1 \mu_3^3} \Pi^2 \wt_2^{1/2}
 \rmk  \,.
\eeq
The maximal value is obtained such as 
\beq
 {\cal P}_\curv^{(\rm peak)}
 &\simeq 
 \frac{\Pi}{2 \sqrt{2\pi} \wt_2 } 
 \lmk 
 1 -   \frac{\Mpl^2}{3 \mu_2^2} \Pi \sqrt{\wt_2} - \frac{3 \Mpl^4}{16 \mu_1 \mu_3^3} \Pi^2 \wt_2
 \rmk
 \nonumber \\
 &\simeq
 \frac{\Ne_\uc}{\sqrt{2 \pi \wt_2^3}} 
\lmk 
 1 -   \frac{ \Mpl^2}{3 \mu_2^2} \Ne_\uc - \frac{ \Mpl^4}{2 \mu_1 \mu_3^3} \Ne_\uc^2
 \rmk
\nonumber \\
&\simeq 0.013 \, \Ne_\uc \lmk \frac{\wt_2}{10} \rmk^{-3/2} \lmk 
 1 -   \frac{ \Mpl^2}{3 \mu_2^2} \Ne_\uc - \frac{ \Mpl^4}{2 \mu_1 \mu_3^3} \Ne_\uc^2
 \rmk  \,,
 \label{Pzeta2}
\eeq
where we use $\wt_2 \gg 1$ and neglect $c/\wt_2 \ll 1$ in the second line. 
The shape of the power spectrum is determined by \eq{width}, 
where 
$\wt_k$ can be written in terms of $\Ne_k (\wt_k)$ by using \eq{N122}.

Now one can see that the effect of the quadratic and cubic terms for $\phi$  
resolves the degeneracy between the peak amplitude and wavenumber. 
From Eqs.~(\ref{eq:NPT}) and (\ref{Pzeta2}), 
the next-to-leading order terms are relevant for $(\Mpl / \mu_2)^2/3 \sim \Ne_\uc$ and/or $\Mpl^4 / (2 \mu_1 \mu_3^3) \sim \Ne_\uc^2$. 
We can then choose $\mu_2$ and $\mu_3$ appropriately to suppress the curvature perturbations for a given $\Pi$ (or $\Ne_\uc$). 
For desired values of $\Ne_\uc$ and $\mathcal{P}_\curv^{\rm (peak)}$, 
the parameters of $\Pi$ and either $\mu_2$ or $\mu_3$ are determined. 
Then we can use the remaining free parameter to make the spectral index consistent with the observed value by \eq{ns}.

There should be a cancellation in the last parenthesis in \eq{Pzeta2} to reduce the amplitude of curvature perturbations. 
One may wonder how much fine-tuning is required to obtain a desired PBH abundance. 
Let us define 
the degrees of fine-tuning such as 
\beq
 \abs{\dv{\ln \mathcal{P}_\curv^{(\rm peak)}}{\ln \mu_3}} 
 \simeq
 -3 \frac{- \frac{ \Mpl^4}{2 \mu_1 \mu_3^3} \Ne_\uc^2}{\lmk 
 1 -   \frac{ \Mpl^2}{3 \mu_2^2} \Ne_\uc - \frac{ \Mpl^4}{2 \mu_1 \mu_3^3} \Ne_\uc^2
 \rmk}  \,, 
 \label{tuning}
\eeq
where we use the analytic result of \eq{Pzeta2}. 
If this quantity is much larger than unity, a fine-tuning for the parameter $\mu_3$ is required to obtain a desired value of $\mathcal{P}_\curv^{(\rm peak)}$. 
For example, 
if $\Pi^2 = 185$, $\mu_2 = 4.21 \Mpl$, and $\mu_3 = 0.182 \Mpl$, which results in 
$\mu_1 / \Mpl \simeq 1.3 \times 10^5$, $\Ne_\uc \simeq 17.3$, 
$\mathcal{P}_\curv^{(\rm peak)} \simeq 0.0141$, and $n_s \simeq 0.969$ from our numerical results shown shortly, 
we obtain the degrees of fine-tuning such as $\abs{{\rm d} \ln \mathcal{P}_\curv^{(\rm peak)} / {\rm d} \ln \mu_3} \simeq 1.2$. 
We therefore conclude that the tuning of the parameter is not severe to reduce the amplitude of curvature perturbations.


\subsection{Numerical results}
\label{sec:numerical}

To check the results of analytical calculations, 
we numerically solve the equation of motion and calculate the curvature perturbations by the $\delta \Ne$ formalism (see \eq{eq:deltaN}). 
We adopt a simplified procedure used in Ref.~\cite{Clesse:2015wea} (see also Ref.~\cite{Kodama:2011vs}), 
where the stochastic dynamics determines an ``initial'' condition of waterfall field $\psi$ at the waterfall phase transition 
and then solve the classical equation of motion without the noise term.

The detail of our numerical calculation is as follows. 
First, we start from the time of the waterfall phase transition, at which $\psi = \psi_0$ and $\phi = \phi_\uc$. The ``initial'' condition of waterfall field $\psi_0$ is determined by the stochastic noise such as \eq{eq:r0}. 
We then solve the classical equation of motion for $\phi$ and $\psi$ without the noise term until the end of inflation. 
From the result, we can calculate the e-folding number at the waterfall phase transition, $\Ne_\uc$. 
By changing the initial condition slightly and again solving the equation of motion, we calculate the peak amplitude of curvature perturbations by the $\delta \mathcal{N}$ formalizm. 
Then we solve the equation of motion for the inflaton $\phi$ 
to the backward-time direction from the waterfall phase transition by using the slow-roll approximation. 
We then calculate the amplitude of curvature perturbations and its spectral index at the pivot scale of the CMB. 
The e-folding number of the pivot scale $\Ne_*$ 
is given by \eq{eq:Nstar} with the assumption of instantaneous reheating (i.e., $H_{\rm RH} = H_{\rm inf}$). 
We also 
require that the total e-folding number is larger than $60$ so that the flatness and horizon problems are addressed by inflation.

\begin{figure}
    \centering
            \includegraphics[width=0.95\hsize]{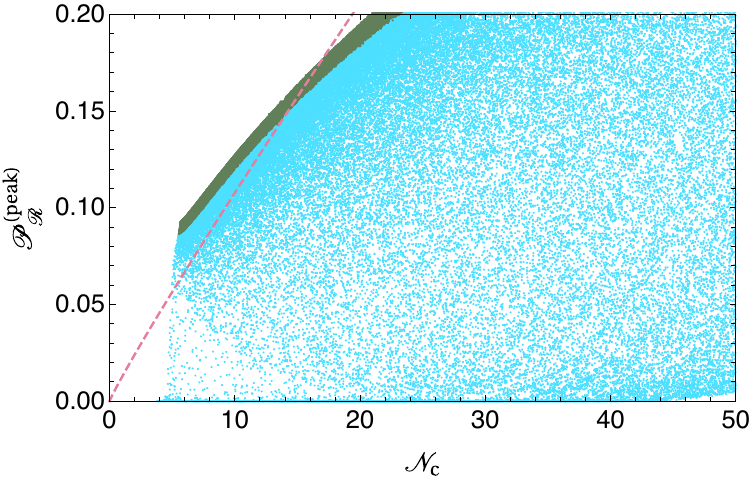}
    \caption{
    The peak amplitude of curvature perturbations $\mathcal{P}_\curv^{(\rm peak)}$ as a function of 
    e-folding number at the waterfall phase transition $\Ne_\uc$.
    The dashed line represents the analytic result of \eq{Pzeta} for the minimal model without quadratic and cubic terms. 
    The green (blue) points represent the results with $\mu_3 / \Mpl \ge 1$ ($\mu_3 / \Mpl < 1$). 
}
    \label{fig:result0}
\end{figure}

We take $\phi_\uc /\sqrt{2} = M = \Mpl/10$ as an example. 
We randomly take the parameters $\Pi$, $\mu_2$, and $\mu_3$ in the following domains: 
\beae{
 &\Pi^2 \in (5, 2000) \,,
 \\
 &\mu_2/\Mpl \in (2, 15) \,,
 \\
 &\mu_3/\Mpl \in (0.03, 10) \,.
}
The parameter $\mu_1$ is determined from $\Pi$. 
These parameter spaces cover the whole parameter space we are interested in. 
Taking those parameters randomly, 
we solve the equation of motion and calculate the curvature perturbations and the spectral index. 
Each point in Fig.~\ref{fig:result0} shows the result of a certain set of parameters in the $\mathcal{P}_\curv^{\rm (peak)}$-$\Ne_\uc$ plane, 
where we keep only the results that are consistent with the observed spectral index $\ns = 0.9649 \pm 0.0042$. 
We plot the results with $\mu_3 / \Mpl \ge 1$ ($\mu_3 / \Mpl < 1$) as green (blue) points. 
The green points represent the case in which the cubic term for the inflaton potential is negligible. 
We can see that $\mathcal{P}_\curv^{\rm (peak)}$ and $\Ne_\uc$ are correlated with each other for green points, 
consistently with the analytic result of \eq{Pzeta} (shown as the red dashed line) for the minimal model without quadratic and cubic terms. 
For a smaller $\mu_3/\Mpl$, 
the degeneracy between $\mathcal{P}_\curv^{\rm (peak)}$ and $\Ne_\uc$ is resolved and a smaller $\mathcal{P}_\curv^{\rm (peak)}$ and larger $\Ne_\uc$ can be realized as shown by blue points.

\begin{figure}
    \centering
            \includegraphics[width=0.95\hsize]{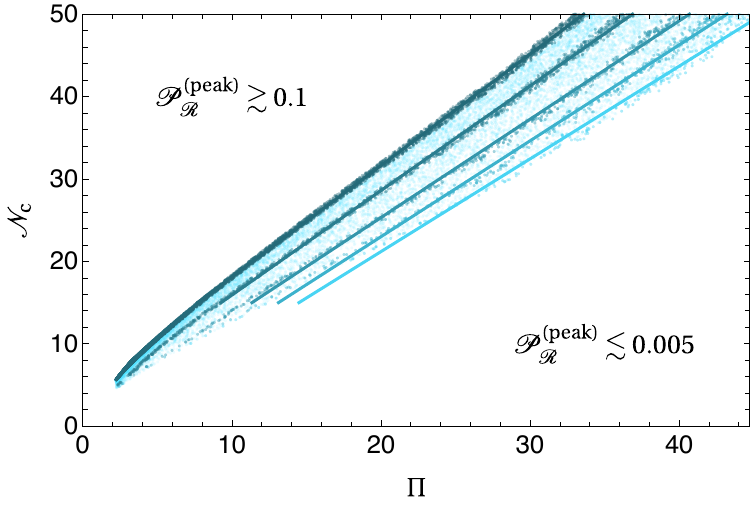}
\\
        \vspace{0.2cm}
            \includegraphics[width=0.95\hsize]{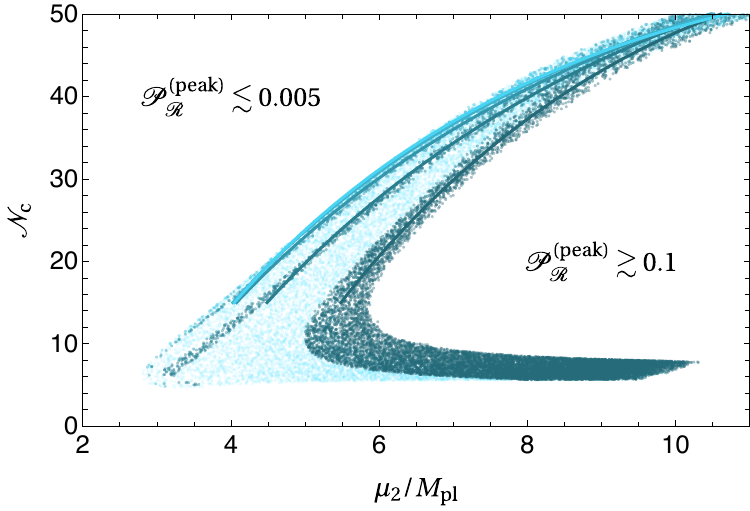}
\\
        \vspace{0.2cm}
            \includegraphics[width=0.95\hsize]{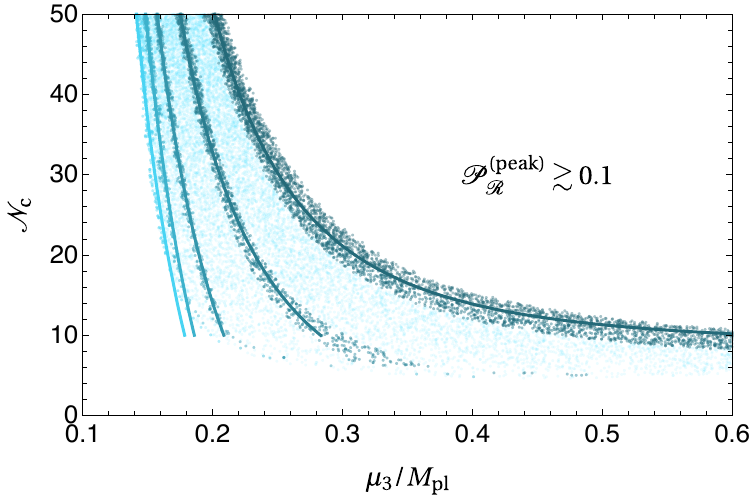}
    \caption{The e-folding number at the waterfall phase transition $\Ne_\uc$ as a function of $\Pi$ (top), $\mu_2/\Mpl$ (middle) $\mu_3/\Mpl$, and (bottom).
    The solid curves represent the fitting functions for $\mathcal{P}_\curv^{(\rm peak)} \simeq 0.1$, $0.05$, $0.02$, $0.01$, and $0.005$ from the dense to light ones. 
    The lightest points include the data for $\mathcal{P}_\curv^{(\rm peak)} \in (0.0045, 0.11)$, whereas the dense points are the ones that are used to determine the fitting curves. 
}
    \label{fig:result1}
\end{figure}

The parameters that satisfy $\ns = 0.9649 \pm 0.0042$ and $\mathcal{P}_\curv^{(\rm peak)} \in (0.0045, 0.11)$ are shown in Fig.~\ref{fig:result1} as light points. 
We plot the points for $\mathcal{P}_\curv^{(\rm peak)} \in (0.9 \mathcal{P}_0, 1.1 \mathcal{P}_0)$ with $\mathcal{P}_0 = 0.005$, $0.01$, $0.02$, $0.05$, and $0.1$ denser for a larger $\mathcal{P}_0$. 
The solid curves represent the fitting functions for $\mathcal{P}_\curv^{(\rm peak)} \simeq \mathcal{P}_0$ with $\mathcal{P}_0 = 0.005$, $0.01$, $0.02$, $0.05$, and $0.1$ from the light to dense ones. 
The fitting functions are given by 
\beq
 &\Ne_\uc \simeq c_{\Pi, 0} + c_{\Pi, 1} \Pi \qquad \text{for} \quad \Ne_\uc > 15 \,, 
 \label{eq:analyticPi}
 \\
 &\Ne_\uc \simeq c_{\mu_2,0} + c_{\mu_2,1/2} \lmk \frac{\mu_2}{\Mpl} \rmk^{1/2} + c_{\mu_2,1} \lmk \frac{\mu_2}{\Mpl} \rmk  \qquad \text{for} \quad \Ne_\uc > 15 \,, 
 \label{eq:analyticmu2}
 \\
 &\Ne_\uc \simeq c_{\mu_3,0} + c_{\mu_3,-3} \lmk \frac{\mu_3}{\Mpl} \rmk^{-3}  \qquad \text{for} \quad \Ne_\uc > 10 \,,
 \label{eq:analyticmu3}
\eeq
for each panel in the figure.
The coefficients are chosen for each $\mathcal{P}_0$ and are given in Table~\ref{tab:1}. 
The form of these functions is determined by the analytic arguments. 
For example, $\Ne_\uc$ is proportional to $\Pi$ as indicated by \eq{eq:NPT}. 
It is also almost independent of $\mu_3 / \Mpl$. This is expected from \eq{Pzeta2} with $\mu_1 \propto \Pi^2 \propto \Ne_\uc^2$, by requiring that the parenthesis should be small to suppress the peak amplitude of curvature perturbations. 
The behavior of $\mu_2 / \Mpl$ is difficult to understand, but its order of magnitude is consistent with the condition to tune the parenthesis in \eq{Pzeta2} and obtain the desired value of spectral index in \eq{ns}.

\begin{figure}
    \centering
            \includegraphics[width=0.95\hsize]{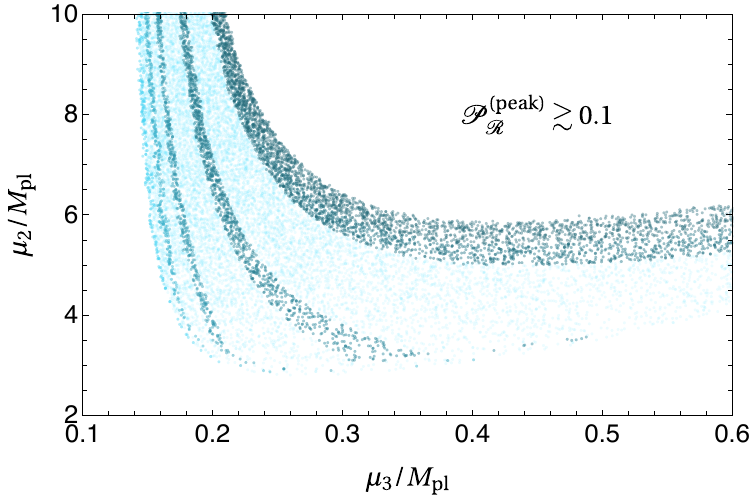}
    \caption{Same as Fig.~\ref{fig:result1} but for $\mu_2/\Mpl$ as a function of $\mu_3/\Mpl$.
}
    \label{fig:result1-2}
\end{figure}

We also plot $\mu_2/\Mpl$-$\mu_3/\Mpl$ plane in Fig.~\ref{fig:result1-2} to clarify the correlation between these parameters. It shows that the quadratic or cubic terms have to be strong enough (i.e., $\mu_2/\Mpl$ or $\mu_3/\Mpl$ have to be small enough) to reduce the amplitude of curvature perturbations.

\begin{table}
\caption{Coefficients for fitting functions.}
\label{tab:1}
\centering
\begin{tabular}{c|rrrrrrr}
 $\mathcal{P}_0$ & \ $c_{\Pi,0}$ & \ $c_{\Pi,1}$ & \ $c_{\mu_2,0}$ & \ $c_{\mu_2,1/2}$ & \ $c_{\mu_2,1}$ & \ $c_{\mu_3,0}$ & \ $c_{\mu_3,-3}$ \\
\hline
 0.1 & 4.6 & 1.4 & -207 & 135 & -17 & 8.6 & 0.33
 \\
 0.05 & 3.2 & 1.3 & -114 & 80 & -9.2 & -2.4 & 0.28
 \\
 0.02 & 1.5 & 1.2 & -95 & 71 & -8.0 & -20 & 0.27
 \\
 0.01 & -0.27 & 1.2 & -103 & 78 & -9.4 & -31 & 0.27
 \\
 0.005 \ \ & \ \ -1.3 & \ \ 1.1 & \ \ -113 & \ \ 86 & \ \ -11 & \ \ -29 & \ \ 0.22
\end{tabular}
\end{table}

Here we comment on the parameter space considered in the literature. 
There are many data with $\Ne_\uc = \mathcal{O}(1)$ for $\mu_2 / \Mpl \gtrsim 6$ as shown at the bottom-right corner in the middle panel in Fig.~\ref{fig:result1}. 
The corresponding points are not shown in the bottom panel because they require $\mu_3 / \Mpl \gtrsim 1$. 
In this parameter space, 
the analytic calculation at the leading order is a good approximation 
and hence they correspond to the case considered in Refs.~\cite{Clesse:2015wea,Kawasaki:2015ppx}. 
However, $\mathcal{P}_\curv^{(\rm peak)}$ cannot be smaller than about $0.1$ in this parameter space. 
A parameter space with a larger $\Ne_\uc$ gives a larger $\mathcal{P}_\curv^{(\rm peak)}$ (see the green points in Fig.~\ref{fig:result0}) . 
This demonstrates that 
astrophysical size PBHs cannot form (or are overproduced) from a hybrid inflation model if the quadratic and cubic terms for the inflaton potential are negligible. 
Including the latter two terms, 
we find the parameter space in which $\Ne_\uc$ is $\mathcal{O}(10)$ and $\mathcal{P}_\curv^{(\rm peak)} \sim 0.01$.

\section{PBH formation}
\label{sec:PBH}

\subsection{Press--Schechter formalism}

Finally, we consider PBH formation from the collapse of overdense regions. 
The large curvature perturbations generated by the stochastic dynamics result in large density perturbations after inflation. 
If the overdensity exceeds a certain threshold, the overdense region tends to collapse to form PBHs with the size corresponding to the mode entering the horizon. 
The perturbation of comoving wavenumber $k$ enters the horizon when $k = a(t) H(t)$. 
The PBH mass is given by the total energy enclosed within the Hubble horizon 
at the horizon crossing: 
\beq
 M_{\rm BH} 
 &= \left. \gamma \rho \frac{4\pi H^{-3}}{3} \right\vert_{ k = a H}, \nonumber \\
 &\simeq 10^{20} \g 
 \lmk \frac{\gamma}{0.2} \rmk 
 \lmk \frac{g_*}{106.75} \rmk^{-1/6} 
 \lmk \frac{k}{7 \times 10^{12} \Mpc^{-1}} \rmk^{-2}  \,,
\eeq
where $\gamma$ ($= {\cal O}(1)$) is a numerical constant~\cite{Carr:1975qj}. 
The corresponding e-folding number at the horizon exit, which we identify $\Ne_\uc$, is given by 
\beq
 \Ne_* - \Ne_\uc 
 &\simeq 32.6
 - \frac{1}{2} \ln \lmk \frac{M_{\rm BH}}{10^{20} \, {\rm g}} \rmk \,, 
 \label{PBH_N_PT}
\eeq
where $\Ne_*$ is given by \eq{eq:Nstar} and we use $\gamma = 0.2$ and $g_* = 106.75$.

We can estimate the PBH abundance by the Press--Schechter theory, assuming that 
the density perturbations are Gaussian 
and that a PBH forms from a density perturbation above a certain threshold $\delta_\uc$ ($\sim 0.3$).%
\footnote{
Note that the estimation scheme for the \ac{PBH} abundance has been intensively developed since the simplest Press--Schechter approach with a uniform threshold $\delta_\uc$. For example, the number density of the overdense region is evaluated by the peak statistics of a random field called \emph{peak theory} (see, e.g., Refs.~\cite{Yoo:2018kvb,Yoo:2019pma,Yoo:2020dkz,Kitajima:2021fpq}). The \ac{PBH} threshold is also derived in a more sophisticated way from the so-called compaction function, the excess of the Misner--Sharp mass from the background one, not uniformly but dependently on the overdensity's profile (see, e.g., Refs.~\cite{Shibata:1999zs,Harada:2015yda,Young:2019yug,Escriva:2019phb,Atal:2019erb}). It is known that the resultant \ac{PBH} mass is not simply given by the horizon mass but features a scaling relation $M\propto(\delta-\delta_\uc)^p$ with a universal power $p\simeq0.36$ (see Refs.~\cite{Choptuik:1992jv,Evans:1994pj,Koike:1995jm,Niemeyer:1997mt,Niemeyer:1999ak,Hawke:2002rf,Musco:2008hv}). We however neglect all these corrections because we do not even know the precise statistics of the curvature perturbation beyond the power spectrum in this model. Even the power spectrum can be modulated by the stochastic effect in the waterfall phase that we neglected.
We therefore leave accurate \ac{PBH} abundance estimation for future works, simply pointing out the resolution of the degeneracy in this paper.}
From these criteria, the probability for the PBH formation is calculated from 
\beq
	\beta (M) 
	&\equiv \frac{\rho_{\rm PBH}(M)}{\rho_{\rm tot}} 
	\approx \int_{\delta_\uc}^\infty 
	\frac{\dd{\delta}}{\sqrt{2 \pi \sigma^2 (M)}} \, e^{- \frac{\delta^2}{2 \sigma^2(M)}}, \nonumber \\
	&\simeq
	\sqrt{\frac{2}{ \pi}} \frac{ \sigma (M)}{\delta_\uc} \, e^{- \frac{\delta_\uc^2}{2 \sigma^2(M)}} \,,
 \label{betaapp} 
\eeq
where $\sigma (M)$ is the variance of the coarse-grained density contrast for the scale of wavenumber $k$ corresponding to the PBH mass $M(k)$~\cite{Young:2014ana}: 
\beq
 \sigma^2 (M(1/R)) = \frac{16}{81} \int \dd \ln k' \lmk k' R \rmk^4 {\cal P}_{\curv} (k') \, W(k' R)^2, \, .
\label{sigmasquared}
\eeq
The window function is taken to be $W (x) = \exp \lmk - x^2 / 2 \rmk$. 
The resulting PBH abundance is given by 
\beq
	f_{\rm PBH} \equiv \frac{\Omega_{\rm PBH}}{\Omega_{\rm DM}^{(\rm obs)}} 
	\simeq 
	\lmk \frac{\beta (M)}{8 \times 10^{-15}} \rmk 
	\lmk \frac{M}{10^{20} \ {\rm g}} \rmk^{-\frac{1}{2}} \,, \,\,\,
	\label{Omega_PBH}
\eeq
where $\Omega_{\rm DM}^{(\rm obs)}$ represents the observed DM abundance and we use $\gamma = 0.2$ and $g_* = 106.75$.

\begin{figure}
    \centering
            \includegraphics[width=0.95\hsize]{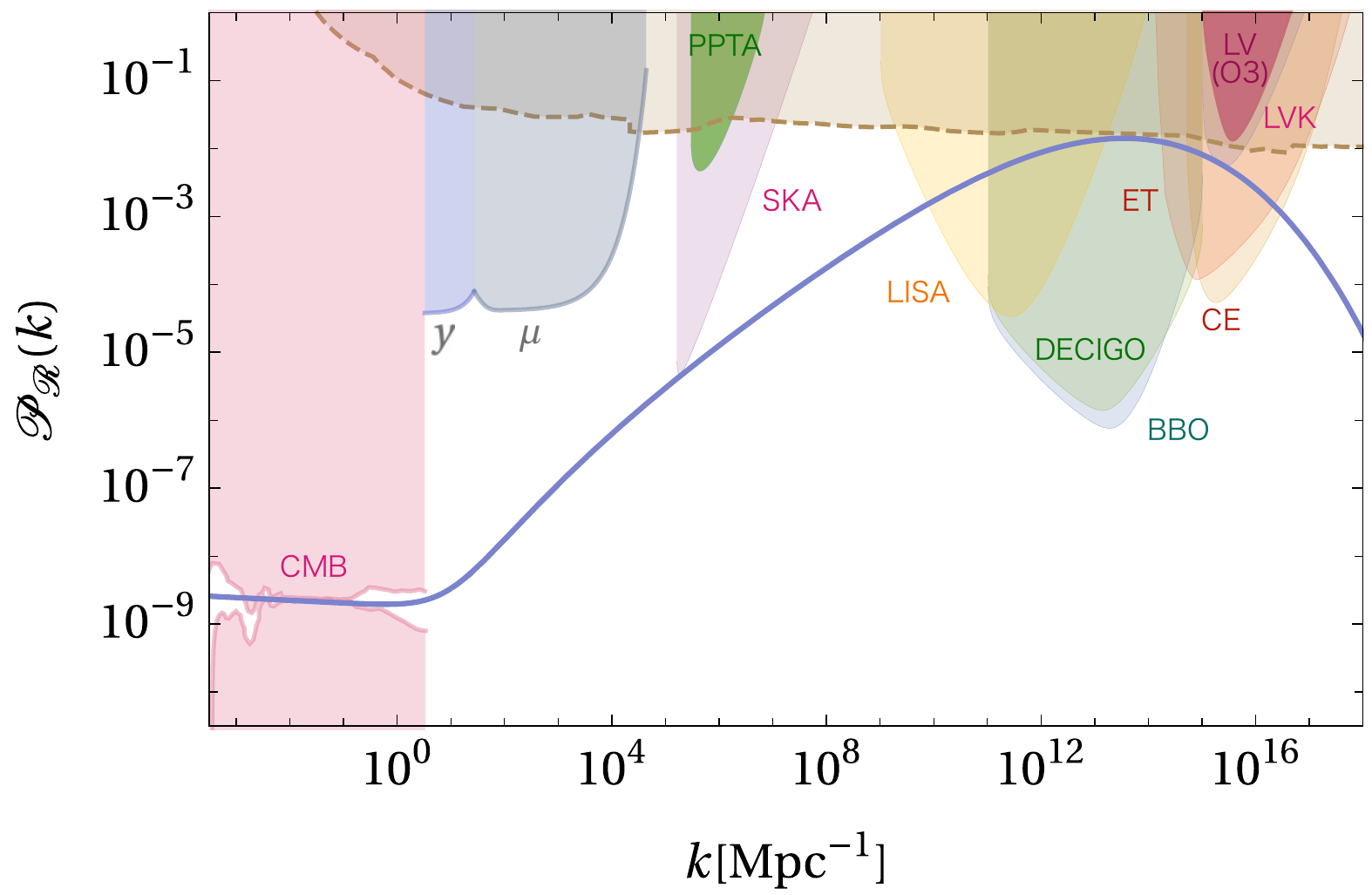}
\\
\vspace{0.2cm}
            \includegraphics[width=0.95\hsize]{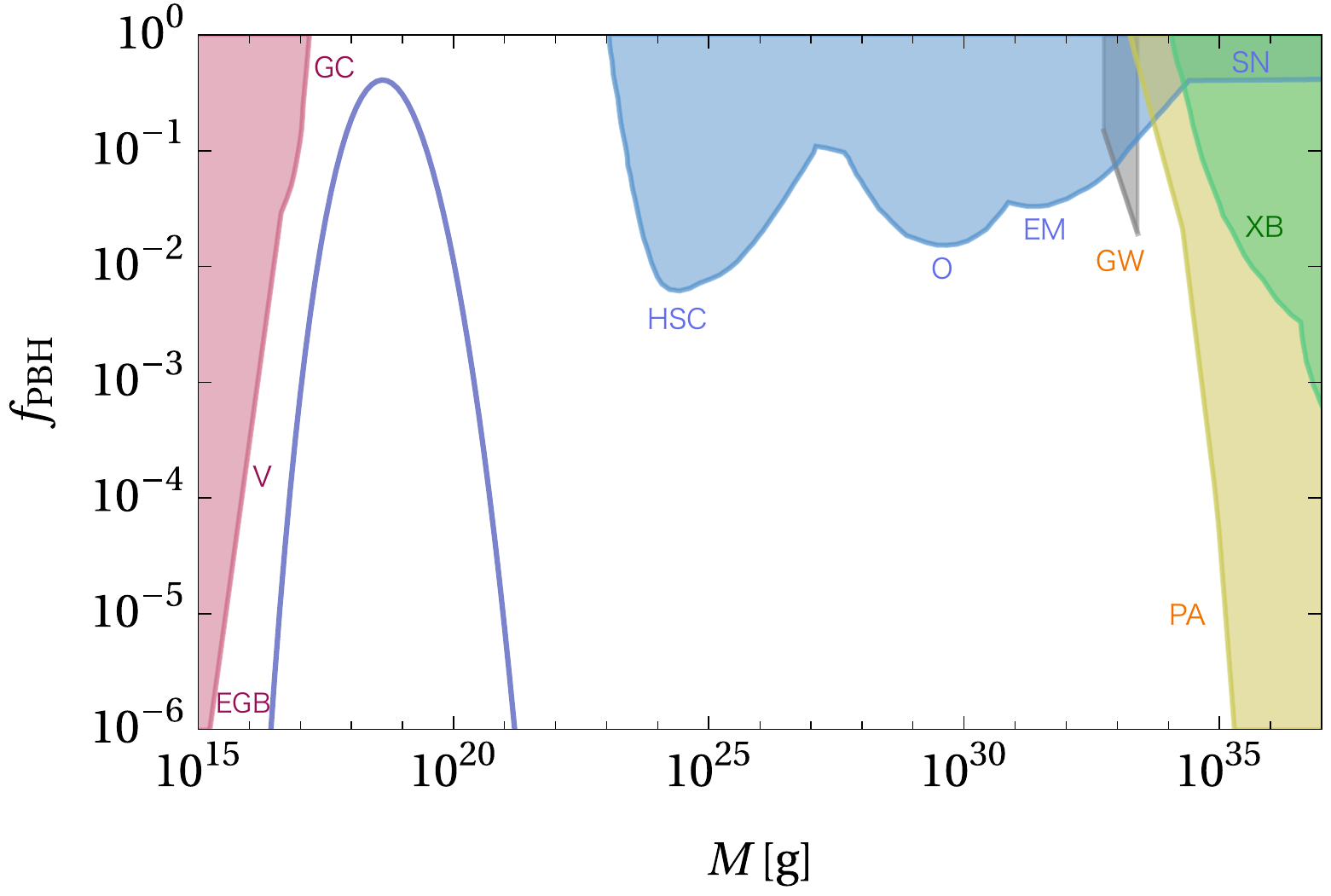}
\\
\vspace{0.2cm}
            \includegraphics[width=0.95\hsize]{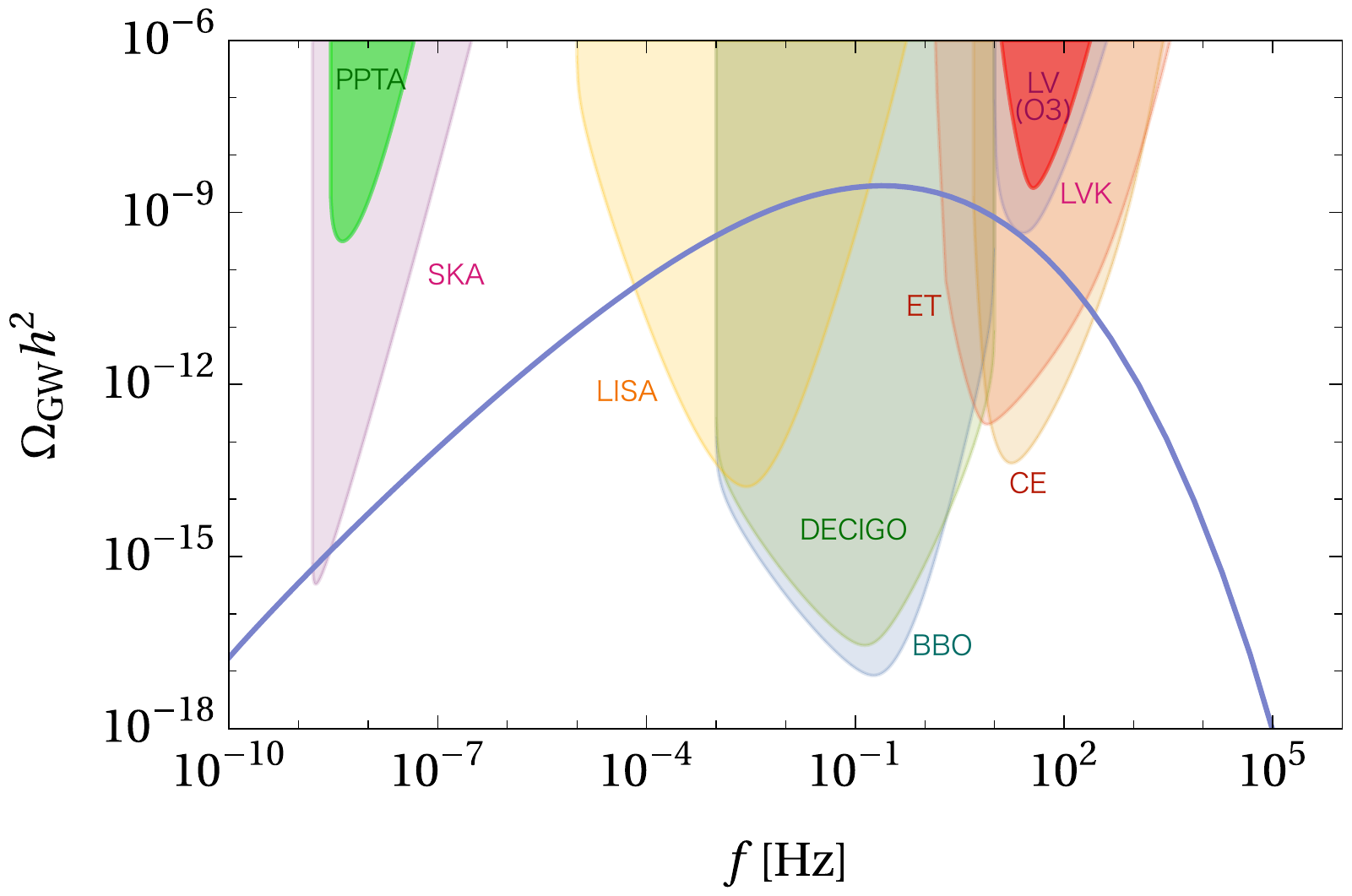}
    \caption{Spectrum of curvature perturbations $\mathcal{P}_\curv$ (upper panel), PBH mass function (middle panel),     
    and gravitational waves $\Omega_{\rm GW}h^2 (f)$ (bottom panel) 
    for the case with $\Pi^2 = 185$, $\mu_2 = 4.21 \Mpl$, and $\mu_3 = 0.182 \Mpl$. The shaded regions represent the constraints and future sensitivity curves (see the main text for detail).  
}
    \label{fig:result2}
\end{figure}

There are several constraints on the PBH abundance for a broad range of the mass scale. The shaded regions in the middle panel in Fig.~\ref{fig:result2} are excluded by evaporation (red), lensing (blue), gravitational waves (gray, GW)~\cite{Raidal:2017mfl}, CMB distortions (orange, PA)~\cite{Serpico:2020ehh}, and accretion from X-ray binaries (green, XB)~\cite{Inoue:2017csr}. 
For more details, 
the evaporation limits come from the extragalactic $\gamma$-ray background (EGB)~\cite{Page:1976wx,Carr:2009jm}, the Voyager positron flux (V)~\cite{Boudaud:2018hqb} and annihilation-line radiation from the Galactic centre (GC)~\cite{Laha:2019ssq, DeRocco:2019fjq}. 
The lensing constraints come from microlensing of supernovae (SN)~\cite{Zumalacarregui:2017qqd} and of stars in M31 by Subaru (HSC)~\cite{Niikura:2017zjd}, the Magellanic Clouds by the Experience pour la Recherche d’Objets Sombres (EROS) and Massive Compact Halo Object (MACHO) collaborations (EM)~\cite{Blaineau:2022nhy}, and the Galactic bulge by the Optical Gravitational Lensing Experiment (OGLE) (O)~\cite{Wyrzykowski:2011tr} (see also Ref.~\cite{Niikura:2019kqi}). 
Here we assume a delta-function spectrum to illustrate the constraints~\cite{Escriva:2022duf}, which should not be compared directly with our nearly-Gaussian spectrum but is useful for illustrative purpose. 
There is a window where PBHs can be all DM, where $M \sim 10^{17 \text{--} 23} \g$ and  $\Omega_{\rm PBH} h^2 \simeq 0.12$.  
Note that this requires $\calP_{\curv} \sim 0.01$ 
from Eqs.~(\ref{betaapp}) and (\ref{Omega_PBH}). 
The corresponding e-folding number $\Ne_\uc$ should satisfy \eq{PBH_N_PT}.

\subsection{PBH formation in hybrid inflation model}

Now we shall consider the PBH formation in the hybrid inflation model. 
Let us first look at the analytic result of Eqs.~(\ref{eq:NPT}) and (\ref{Pzeta2}) for $\mathcal{P}_\curv$ and $\Ne_\uc$. 
If we omit the corrections from the quadratic and cubic potentials for the inflaton, 
which corresponds to the second and third terms in the parenthesis in \eq{Pzeta2}, 
the peak amplitude is as large as $0.2$--$0.3$ for $\Ne_{\rm PT} \simeq 21$. This results in an overproduction of PBHs by many orders of magnitude. 
However, the peak amplitude can be smaller thanks to the corrections from the quadratic and cubic terms. 
This is the main idea of the present paper 
and the corresponding parameter space is demonstrated as the light blue curves in Fig.~\ref{fig:result1}.

The upper panel in Fig.~\ref{fig:result2} shows the spectrum of curvature perturbations \eq{width} with ${\cal P}_\curv^{(\rm peak)}$ obtained by numerical calculation and $\chi_k$ given by the analytic one \eq{N122} for the case with $\Pi^2 = 185$, $\mu_2 = 4.21 \Mpl$, and $\mu_3 = 0.182 \Mpl$, which results in $\Ne_\uc \simeq 17.3$, $\mathcal{P}_\curv^{(\rm peak)} \simeq 0.0141$, and $n_s \simeq 0.969$. 
The corresponding PBH mass function is shown as the solid curve in the middle panel, 
where the peak PBH mass is $5.3 \times 10^{18} \, {\rm g}$ and the total PBH abundance is $\Omega_{\rm PBH} h^2 \simeq 0.1$. 
The shaded regions in the upper panel represent the excluded regions or future sensitivity curves. 
A parameter space with a relatively small wavenumber is excluded 
by constraints of CMB temperature anisotropies~\cite{Nicholson:2009pi, Nicholson:2009zj, Bird:2010mp, Bringmann:2011ut} and $\mu$- and $y$-distortions~\cite{Fixsen:1996nj, Chluba:2012we}.%
\footnote{See also Ref.~\cite{Bianchini:2022dqh} for recent analysis, which improved the constraint for $\mu$-distortion by a factor of $2$.} 
The upper-most region, denoted by the dotted curve, is excluded by the overproduction of PBHs (that partially corresponds to the shaded regions in the middle panel in Fig.~\ref{fig:result2}), where we adopt the constraint for Gaussian spectrum by referring to Fig.~19 in Ref.~\cite{Carr:2020gox}.

Large curvature perturbations can generate the stochastic \ac{GW} background through the second-order effect~\cite{Saito:2008jc, Saito:2009jt}. 
We calculate the GW spectrum in hybrid inflation model and compare the result with constraints and future sensitivity curves for the GW experiments. 
The bottom panel in Fig.~\ref{fig:result2} shows the prediction of GW spectrum 
for the above-mentioned parameter. 
It is compared with the power-law-integrated sensitivity curves for ongoing and planned GW experiments~\cite{Schmitz:2020syl}. 
Pulsar-timing array (PTA) experiments, such as Parkes Pulsar Timing Array (PPTA)~\cite{Shannon:2015ect} exclude the dense green shaded region. 
The aLIGO/aVirgo's third observing run (LV(O3))~\cite{KAGRA:2021kbb} excludes the dense red shaded region.%
\footnote{See also Refs.~\cite{Kapadia:2020pnr,Romero-Rodriguez:2021aws} for a specific analysis of the constraint on the induced \acp{GW}.} 
We also plot future sensitivity curves for PTA and GW detection experiments by light shaded regions, 
including 
SKA~\cite{Janssen:2014dka},
LISA~\cite{LISA:2017pwj},
DECIGO~\cite{Kawamura:2011zz,Kawamura:2020pcg},
BBO~\cite{Harry:2006fi},
Einstein Telescope (ET)~\cite{Punturo:2010zz,Maggiore:2019uih},
Cosmic Explorer (CE)~\cite{Reitze:2019iox},
and aLIGO+aVirgo+KAGRA (LVK)~\cite{Somiya:2011np,KAGRA:2020cvd}. 
The upper panel in Fig.~\ref{fig:result2} also shows them in terms of the curvature perturbations.

The case with the solid curve can be tested by future GW experiments, such as LISA. 
This is an interesting smoking gun signal for PBH formation in the hybrid inflation model. 
In particular, our GW spectrum has a relatively broad peak, which comes from the spectrum of the density curvature perturbations \eq{width}. 
If we can determine the GW spectrum by those experiments, we can obtain information of parameters for the hybrid inflation model, such as $\Pi$, $\mu_2$, and $\mu_3$. 
Then one can check the consistency with the parameter space shown in Fig.~\ref{fig:result1}. 
One can therefore falsify or confirm our model by observations of the GW spectrum, which is actually within the future sensitivity curve.


\section{Discussion and conclusions}
\label{sec:conclusions}

We have revisited the spectrum of curvature perturbations generated during the waterfall phase transition in a hybrid inflation model. After emphasizing the fact that the peak amplitude and wavenumber are correlated with each other in a minimal setup, we have pointed out that their degeneracy can be relaxed by quadratic and cubic terms for the inflaton potential that are omitted in the literature. 
In particular, PBHs with masses of the order of $10^{17\text{--}23}\,\si{g}$ can be generated consistently with any existence constraints. Moreover, the scenario can be tested by observing GW signals induced by second-order scalar perturbations. 

The peak amplitude of curvature perturbations can be reduced by the effect of quadratic and cubic terms in the inflaton potential. This requires tuning in a parameter, which we quantify as \eq{tuning}. The amount of tuning is, however, not that large because we need to reduce the peak amplitude only by a factor of about 10 in order not to overproduce PBHs. 

Still, one has to take particular care of the tail of the spectrum for curvature perturbations. As one can see in Fig.~\ref{fig:result2} or the analytic formula \eq{width} with \eq{N122}, the spectrum generated during the waterfall phase transition is not sharp but is widely distributed over many orders of magnitude in the wavenumber. In particular, it should not affect the spectrum around the CMB scale. This constrains $\Ne_\uc \lesssim 17$ for the parameters we chose throughout this paper. One can therefore generate a PBH with mass of the order of $10^{17\text{--}19}\,\si{g}$ but cannot generate arbitrarily larger PBHs. 
However, this condition can be relaxed if one considers a relatively low inflation energy scale and/or low reheating temperature scenario.

Other than quadratic or cubic terms in the inflaton potential, a multiple number of waterfall fields~\cite{Halpern:2014mca,Tada:2023fvd}, a small explicit breaking of symmetry in the waterfall potential~\cite{Braglia:2022phb}, etc. can be a resolution of the $\calP_\curv$-$\calN_\uc$ degeneracy. 
Those scenarios are motivated by avoiding the domain-wall problem in the simplest $Z_2$ symmetric model, where the domain walls are produced after the waterfall phase transition and then make the Universe highly inhomogeneous by dominating the energy density. In the present paper, we implicitly assumed either of these extensions to avoid the problem while their effect is negligible on the curvature perturbations. 
The non-Gaussian effect on \ac{PBH} abundance (see, e.g., Refs.~\cite{Bullock:1996at,Ivanov:1997ia,Yokoyama:1998pt,Hidalgo:2007vk,Byrnes:2012yx,Bugaev:2013vba,Young:2015cyn,Nakama:2016gzw,Ando:2017veq,Franciolini:2018vbk,Atal:2018neu,Passaglia:2018ixg,Atal:2019cdz,Atal:2019erb,Yoo:2019pma,Taoso:2021uvl,Kitajima:2021fpq,Escriva:2022pnz}), its clustering (see, e.g., Refs.~\cite{Chisholm:2005vm,Young:2014ana,Young:2014oea,Tada:2015noa,Young:2015kda,Suyama:2019cst,Young:2019gfc}), and the induced \acp{GW} (see, e.g., Refs.~\cite{Cai:2018dig,Unal:2018yaa,Yuan:2020iwf,Atal:2021jyo,Adshead:2021hnm,Garcia-Saenz:2022tzu,Abe:2022xur} and Ref.~\cite{Domenech:2021ztg} for a recent review) is also an interesting topic as the perturbation generated through the waterfall transition is expected to show non-vanishing non-Gaussianity~\cite{Kawasaki:2015ppx}.
Recently, the quantum loop correction from the \ac{PBH}-scale perturbation to the \ac{CMB}-scale one is attracting much attention~\cite{Inomata:2022yte,Kristiano:2022maq,Riotto:2023hoz,Choudhury:2023vuj,Choudhury:2023jlt,Kristiano:2023scm,Riotto:2023gpm,Choudhury:2023rks,Firouzjahi:2023aum,Motohashi:2023syh}. Though our model is not a single-field model which conflicts with the hypothetical No-Go theorem by Ref.~\cite{Kristiano:2022maq}, the loop correction in this model would be worth considering.
We leave them for future works.



\acknowledgments

Y.T. is supported by JSPS KAKENHI, Grants 
No. JP19K14707 and No. JP21K13918.
MY was supported by MEXT Leading Initiative for Excellent Young Researchers, and JSPS KAKENHI. Grant Nos.\ JP20H0585, JP21K13910, and JP23K13092.


\bibliographystyle{JHEP}
\bibliography{letter}


\end{document}